\documentclass[sigconf]{acmart}

\copyrightyear{2023}
\acmYear{2023}
\setcopyright{acmlicensed}\acmConference[MM '23]{Proceedings of the 31st
ACM International Conference on Multimedia}{October 29-November 3,
2023}{Ottawa, ON, Canada}
\acmBooktitle{Proceedings of the 31st ACM International Conference on
Multimedia (MM '23), October 29-November 3, 2023, Ottawa, ON, Canada}
\acmPrice{15.00}
\acmDOI{10.1145/3581783.3612134}
\acmISBN{979-8-4007-0108-5/23/10}

\usepackage{enumitem}
\usepackage{multirow}
\usepackage{pifont}
\usepackage{bbding}
\usepackage{hyperref}
\usepackage{makecell}
\usepackage[ruled,vlined]{algorithm2e}

\acmSubmissionID{1921}

\def\ie{$i.e.$}
\def\eg{$e.g.$}

\begin{document}

\title{Cuing Without Sharing: A Federated Cued Speech Recognition Framework via Mutual Knowledge Distillation}

\author{Yuxuan Zhang}
\authornote{Both authors contributed equally to this research.}
\affiliation{%
  \institution{SRIBD, The Chinese University of Hong Kong, Shenzhen}
  \city{Shenzhen}
  \country{China}
}
\email{221041056@link.cuhk.edu.cn}

\author{Lei Liu}
\authornotemark[1]
\affiliation{%
  \institution{SRIBD, The Chinese University of Hong Kong, Shenzhen}
  \city{Shenzhen}
  \country{China}
  }
\email{leiliu@link.cuhk.edu.cn}

\author{Li Liu}
\authornote{Corresponding author.}
\affiliation{%
  \institution{The Hong Kong University of Science and Technology (Guangzhou)}
 \city{Guangzhou}
 \country{China}
}
\email{avrillliu@hkust-gz.edu.cn}








\begin{abstract}
Cued Speech (CS) is a visual coding tool to encode spoken languages at the phonetic level, which combines lip-reading and hand gestures to effectively assist communication among people with hearing impairments. The Automatic CS Recognition (ACSR) task aims to recognize CS videos into linguistic texts, which involves both lips and hands as two distinct modalities conveying complementary information. However, the traditional centralized training approach poses potential privacy risks due to the use of facial and gesture videos in CS data. To address this issue, we propose a new \textbf{Fed}erated \textbf{C}ued \textbf{S}peech \textbf{R}ecognition (FedCSR) framework to train an ACSR model over the decentralized CS data without sharing private information. In particular, a mutual knowledge distillation method is proposed to maintain cross-modal semantic consistency of the Non-IID CS data, which ensures learning a unified feature space for both linguistic and visual information. On the server side, a globally shared linguistic model is trained to capture the long-term dependencies in the text sentences, which is aligned with the visual information from the local clients via visual-to-linguistic distillation. On the client side, the visual model of each client is trained with its own local data, assisted by linguistic-to-visual distillation treating the linguistic model as the teacher. To the best of our knowledge, this is the first approach to consider the federated ACSR task for privacy protection. Experimental results on the Chinese CS dataset with multiple cuers\footnote{The speaker to perform CS coding system for communications is called cuer.} demonstrate that our approach outperforms both mainstream federated learning baselines and existing centralized state-of-the-art ACSR methods, achieving 9.7\% performance improvement for character error rate (CER) and 15.0\% for word error rate (WER). Code is available at \href{https://github.com/YuxuanZHANG0713/FedCSR}{https://github.com/YuxuanZHANG0713/FedCSR}.
\end{abstract}





\maketitle

\section{Introduction}
\begin{figure}[!t]
\centering
\includegraphics[width=0.95\linewidth]{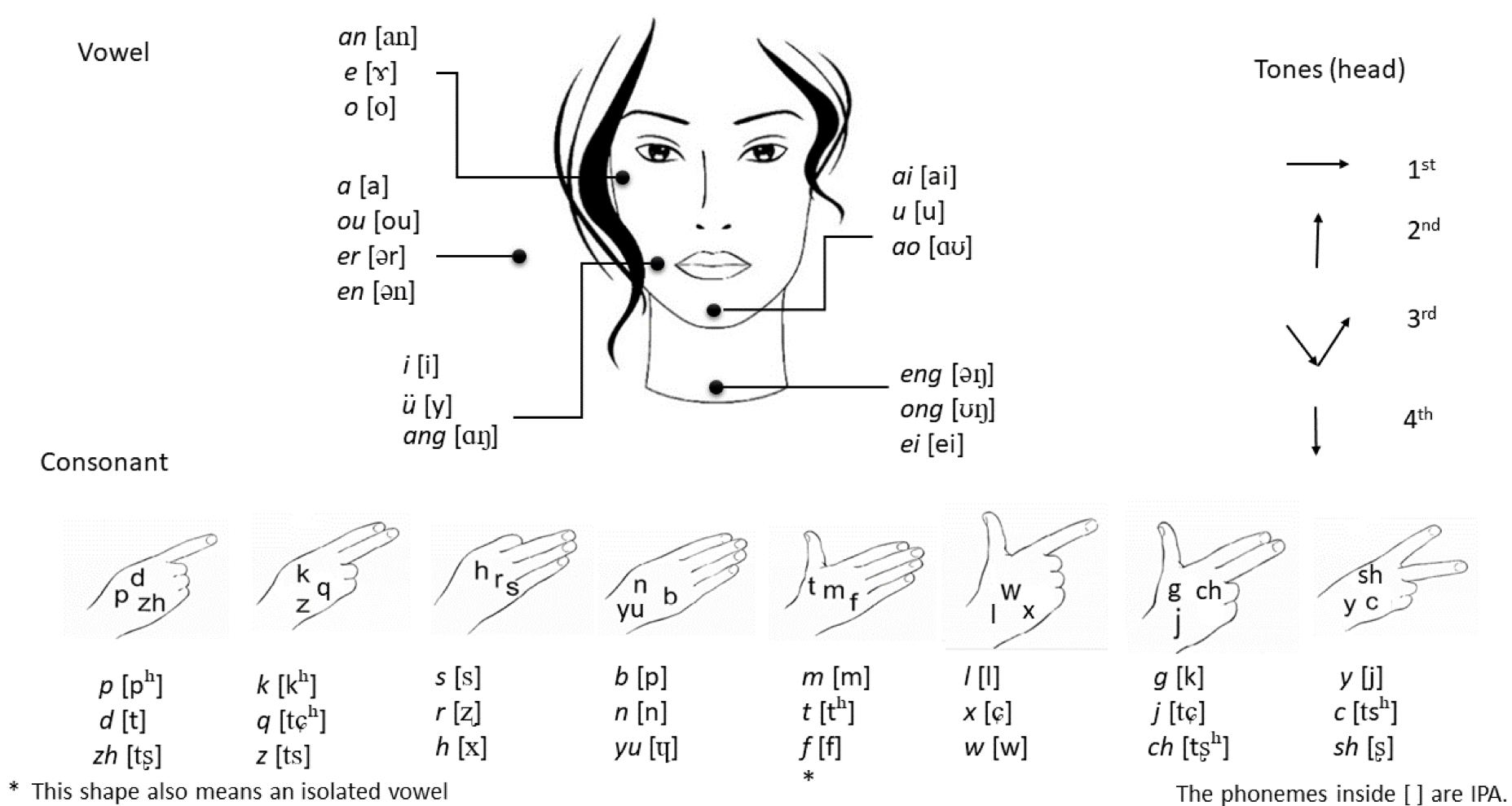}
\caption{The chart for the Mandarin Chinese Cued Speech (MCCS, from \cite{liu2019pilot}), where five different hand positions are used to encode vowels, and eight hand shapes are used to encode consonants in Mandarin Chinese.}
\label{chineseCS}
\vspace{-0.2cm}
\end{figure}

According to the official reports of the World Health Organization (WHO) and the National Deaf Children's Society, over 5\% of the world's population (466 million people) suffer from hearing loss, which influences their daily communications. Lip reading \cite{lip1,liu2017inner} is the most common communication way for hearing-impaired people, which enables them to access spoken speech to improve their communication ability. However, lip reading is easily influenced by the confusing pronunciation, \textit{e.g.}, distinguishing the similar labial shapes of \textit{[u]} and \textit{[y]}, making it difficult to effectively represent linguistic texts. To tackle the disadvantage of lip reading, Cued Speech (CS) \cite{cornett1967cued} is designed to utilize hand movements as the complementary information of lips (see in Figure \ref{chineseCS} for Mandarin Chinese CS as an example), which can provide clear visibility of all phonemes in a spoken language \cite{lip1,wang2021self} and further improve the reading ability of hearing-impaired people.

Considering the large population of hearing-impaired people worldwide, existing researches gradually focus on the Automatic CS Recognition (ACSR) task to improve their communication efficiency. Previous ACSR studies mainly concentrate on the centralized data setting, which aims to learn a feature space with the discriminative ability \cite{liu2018visual}. For example, \cite{heracleous2012continuous, wang2021self} utilized artificial visual marks to capture the features of the regions of interests (ROIs). Speech information to augment visual data via knowledge distillation was exploited in \cite{wang2021cross}. In \cite{liu2020re,gao2023novel}, a re-synchronization procedure was proposed to handle asynchronous modalities in the ACSR task. Besides, considering context relationships in long-time CS videos, \cite{liu2022cross} proposed a cross-modal mutual learning framework, which applies the transformer architecture \cite{vaswani2017attention} to enhance the interaction of multi-modal sequences for the ACSR task. However, the aforementioned approaches directly trained an ACSR model over the CS data of different cuers, which may raise critical privacy issues, \textit{e.g.}, cuer profile leakage due to the facial and gesture images of the cuers in the CS data. Therefore, it is necessary to develop privacy-preserving ACSR algorithms while obtaining public trust for real-world applications. 

A straightforward solution is applying federated learning (FL) to train a global model over decentralized data. The typical FL paradigm is Federated Averaging (FedAvg) \cite{mcmahan2017communication} consisting of the following steps. Firstly, clients download the model from a server. After performing local training using local data, clients upload accumulated gradients to the server. Then the server aggregates the gradients via averaging. Finally, the server broadcasts the updated model to all clients. The FedAvg framework can aggregate the updated gradients of local models without sharing local data, meeting the essential requirements of privacy protection and data security. However, when directly applying such FL framework to the ACSR task, one of the most significant challenges is data heterogeneity caused by Non-IID CS data \cite{ZHU2021noniid,wei2023fedads}, which easily leads to deteriorated convergence in the training process and performs poor recognition accuracy for the ACSR task. 

\begin{figure}[!t]
\centering
\includegraphics[width=0.79\linewidth]{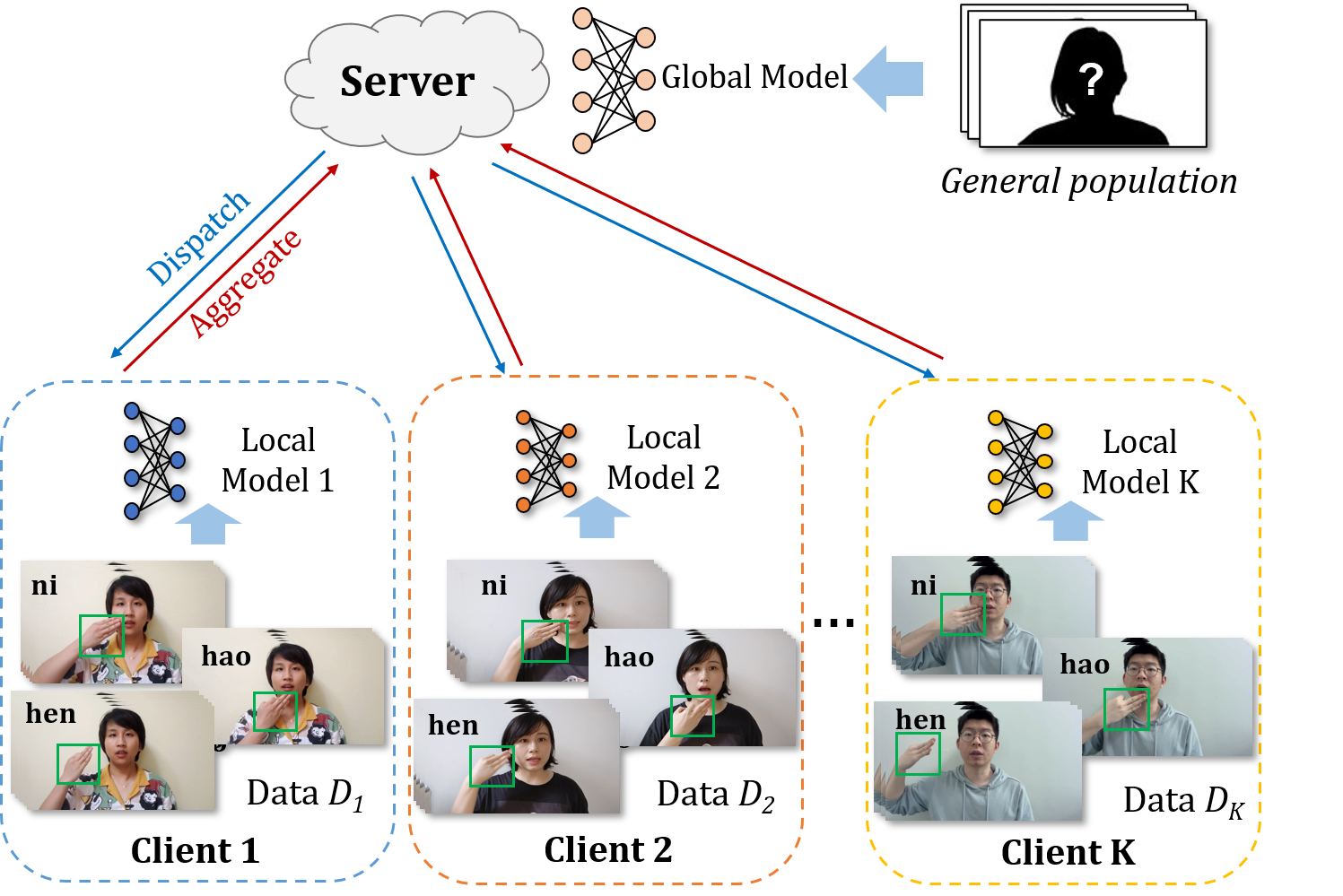}
\caption{Illustration of the ACSR task under the FL setting. Different clients contain CS data of different cuers, indicating the data heterogeneity issue across clients, \textit{e.g.}, three cuers shows different hand position when cuing \textit{[hen]}, and different lip shape when pronouncing \textit{[ni]}. }
\vspace{-0.6cm}
\label{general}
\end{figure}

Two phenomenons are observed that cause the data heterogeneity in the ACSR task under the FL setting. \textbf{The first is from the disturbance of the domain(cuer)-specific information.} As shown in Figure \ref{general}, cuer-specific features (\textit{e.g.}, background and appearance of a cuer) occupy the most image region, while ROIs of lips and hands are concentrated on the small image areas. Such cuer-specific features are different for distinct cuers due to their individual environments. Besides, cuers have their own habits of using the same CS system. For example, when all cuers give the CS representation of \textit{[ren]}, their movements may deviate from the standard hand shapes and positions. Their fluency also results in different speeds and rhythms of speech. \textbf{The second is the natural modality asynchrony between lip and hand movements in CS.} The hand generally moves faster than lips to prepare the next phoneme in a CS system, called hand preceding phenomenon \cite{liu2020re,liu2022objective}. Ideally, the training process should map different cuers' multi-modal data to a unified feature space with semantic consistency. However, under the FL setting, clients may learn inconsistent cross-modal semantic information, \textit{i.e.}, multiple modalities are aligned in a single client but are misaligned across different clients. Therefore, for the ACSR task under the FL setting, it is essential to investigate how to learn domain(cuer)-invariant features with cross-modal semantic consistency, alleviating the negative effects caused by cuer-specific information and multi-modal asynchrony.

To overcome the above challenges, we propose a new \textbf{Fed}erated \textbf{C}ued \textbf{S}peech \textbf{R}ecognition (FedCSR) framework to learn an ACSR model over the decentralized CS data with preserving the cuer privacy. The core idea is to focus on the semantic properties of the CS data to learn cuer-invariant linguistic features. Multi-modal alignment towards the cuer-invariant linguistic features can enhance the cross-modal semantic consistency. Specifically, we propose a mutual knowledge distillation to learn a unified space between linguistic and visual information. Concretely, the server learns a lightweight linguistic model to capture the long-term dependencies in the text sentences, which is constrained by visual-to-linguistic (Vis2Lin) distillation treating the aggregation of local visual models as a teacher. For local training, the visual model of each client is trained under the linguistic-to-visual (Lin2Vis) 
distillation by taking the linguistic model as a teacher. 
Besides, the embedding layers of both visual and linguistic models are globally shared for the consistent embedding space.

The main contributions of this work are summarized as follows:
\begin{itemize}
    \item To alleviate the potential privacy risks caused by the personal information in the CS data, we propose a new framework called FedCSR to learn an ACSR model over the decentralized CS data without sharing privacy knowledge. To the best of our knowledge, this is the first work to consider the federated ACSR task for privacy protection. 
    \item We propose a mutual knowledge distillation method to maintain the cross-modal semantic consistency of the Non-IID CS data, which ensures learning a unified feature space for the linguistic and visual information. 
    \item The experimental results on the Chinese CS dataset with multiple cuers demonstrate that our approach outperforms not only mainstream FL baselines, but also existing centralized state-of-the-art ACSR methods by 9.7\% lower character error rate (CER) and 15.0\% lower word error rate (WER). Extensive ablation studies show that the proposed method can learn a unified feature space for Non-IID CS data.
\end{itemize}

\section{Related Works}
\subsection{Automatic Cued Speech Recognition}
ACSR aims at the recognition task from multiple modalities extracted from a video sequence to a text sentence \cite{liu2018automatic}. Early ACSR studies mainly focus on the recognition of isolated phonemes \cite{heracleous2010cued}, whereas most following works are for decoding continuous CS sentences \cite{heracleous2012continuous}. Recent works generally extract hand and lip features at first and then perform recognition on these modalities, which have become a mainstream paradigm of ACSR. \cite{liu2018visual} extracted the visual features using convolutional neural networks (CNNs) and model the dynamic using Hidden Markov Models (HMMs). Their following work \cite{8903053} proposed a new multi-modal fusion approach for re-synchronization to address the asynchrony issues. \cite{9287365} proposed a fully convolutional architecture, where the features of lips, hand shapes, and hand position are extracted by CNNs. Then these features are concatenated and fed into a time-depth-separable (TDS) encoder structure, followed by a multi-step attention-based phoneme prediction convolution decoder. \cite{wang2021cross} used knowledge distillation from audio data for teaching feature learning on the CS data, alleviating the limited data in CS. \cite{wang2021self} further explored self-supervised contrastive learning for CS representations. \cite{multistream} achieved continuous decoding of CS using pre-trained feature extractor and multi-stream RNN. The most recent work \cite{liu2022cross} presents a multi-modal transformer to extract linguistic and visual features simultaneously and achieves state-of-the-art performance on multiple CS datasets. However, the Non-IID data of different cuers in Chinese CS dataset caused a severe performance drop when training with data of multiple cuers.

\begin{figure*}[!t]
    \centering
    \includegraphics[width=0.9\textwidth]{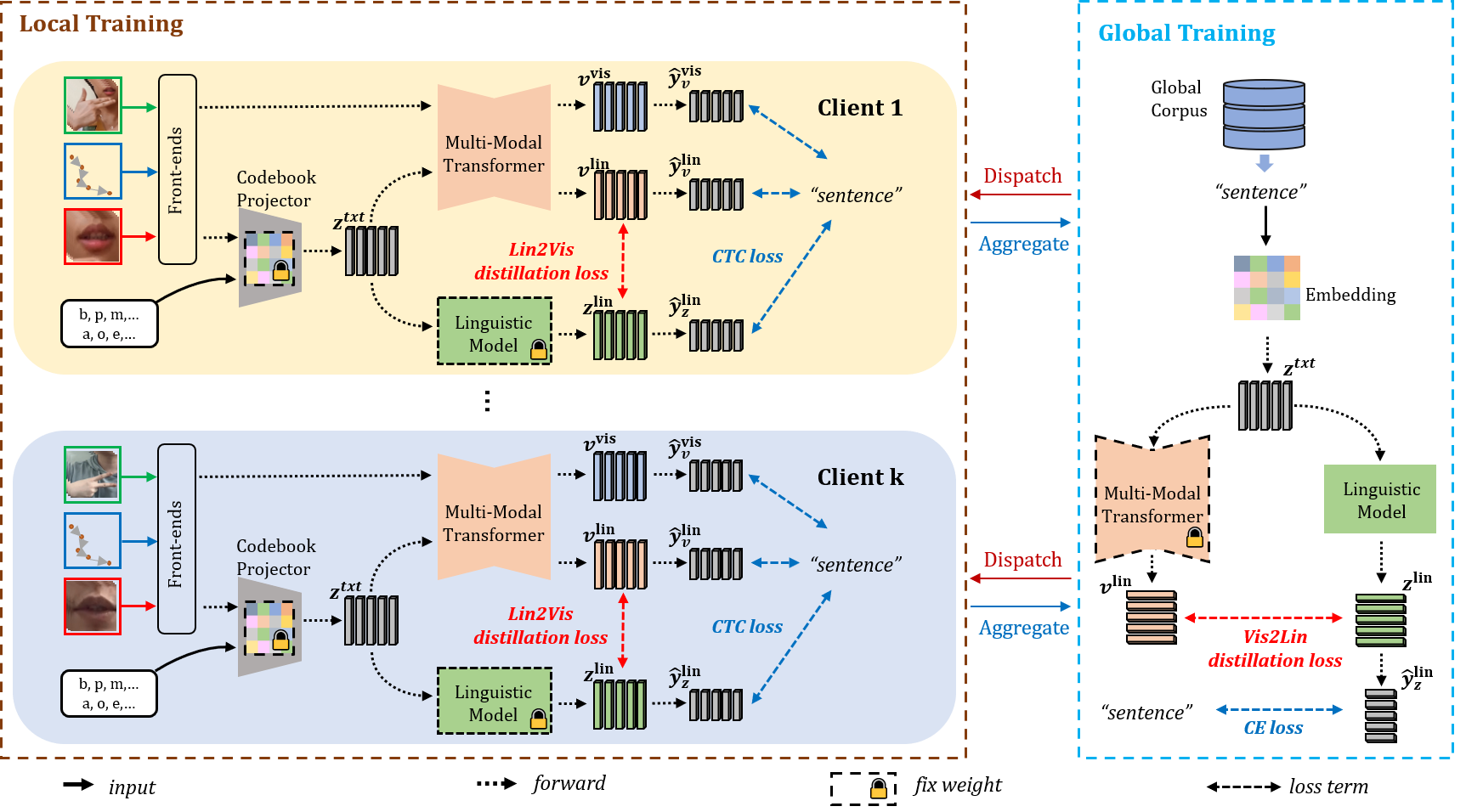}
    \caption{Illustration of the FedCSR framework. The main steps include: (1) Clients download the linguistic and visual model from the server; (2) During local training, the linguistic model serves as the teacher to train a visual model; (3) Clients upload the accumulated gradients of the visual model to the server; (4) The server aggregates the accumulated gradients of the visual model via federated averaging. Then the aggregated visual models serve as the teacher to train the linguistic model.}
    \label{fig:2}
\end{figure*}

\subsection{Federated Learning}
As a technique for learning intelligent models over decentralized private data, FL can be effectively integrated with deep neural networks. The original form of FedAvg \cite{mcmahan2017communication} aggregates the accumulated gradients collected from different clients. During the training process, the training data is always kept on the local devices to protect the privacy of participant data, which also makes communication and computing costs controllable. 
Due to increasing privacy concerns, the utilization of the FL algorithm has grown in various research fields, including health care \cite{med1, med2, med3} and person re-identification \cite{reid1, reid2, reid3}. Combining FL with large-scale video data is more challenging. \cite{rehman2022videossl} explored self-supervised learning for video representation using multiple aggregation strategies and partial weight updating. \cite{feelingwosharing} proposed hybrid aggregation FL method through client clustering for video emotion recognition tasks.

In a realistic scenario, data on different client devices are usually Not Independent and Identically Distributed (Non-IID), which leads to poor performance and lower convergence speed of the aggregated global model. To solve the Non-IID problem, many optimized FL algorithms are proposed in recent years. FedProx \cite{li2020federated} introduced a proximal term and an inexact minimizer that dynamically adjusts local epochs in the local training process, so that local updates are constrained to be closer to the global modal. SCAFFOLD \cite{karimireddy2020scaffold} proposed a strategy to modify the local model that prevents drifted local updates. MOON \cite{Li_2021_CVPR} used model-level federated contrastive learning with model contrastive loss to alleviate the Non-IID issue. The above methods mainly focus on scenarios where clients have different label distributions. FedBN \cite{li2021fedbn} alleviated the impact of feature shift by adding local batch normalization layers that do not participate in aggregation. These works studied the federated optimizer for the general problem where experiments are conducted on simple image classification datasets. Since the Non-IID problem is common among FL applications\cite{ZHU2021noniid}, designing of FL paradigm for specific tasks is necessary.

\section{Preliminaries}
\subsection{Problem Setup}
We discuss a typical FL setting for the ACSR task, \textit{i.e.}, a decentralized system where the CS data is distributed on $K$ client devices. Assume each client is drawn from one of the distinct data sources $\mathcal{D}_1, \mathcal{D}_2, \cdots,\mathcal{D}_K$. The ACSR model $f_w$ parameterized by $w$ is mapping multi-modal data stream (\textit{i.e.}, lip motion, hand gesture, and hand position) into the text, where lip and hand sequences are complementary to each other as distinct modalities. To be simplified, we denote $m\in \{l, h\}$ as lip ($l$) and hand ($h$) modalities in the following section, respectively. The target is to learn an accurate ACSR model that achieves higher recognition performance with protected user profile information. The basic paradigm is to train a shared global model without access to clients' data, which minimizes the following objective function:
\begin{equation}
\small
\min _{w} \frac{1}{K} \sum_{k=1}^K L_{k}(w), 
 \, \textit { where } 
 \, L_{k}(w)=\frac{1}{n_k} \sum_{(\boldsymbol{x}, y) \in \mathcal{D}_k} l_{w}(\boldsymbol{x}, y),
\end{equation}
where $l_w$ is the training loss function and $n_k$ is the sample number of $\mathcal{D}_k$. In the presence of heterogeneity over $\{\mathcal{D}_k\}$, the above target suffers from inconsistent semantic space across clients. Thus, we consider $\min_{w} \mathbb{E}_{(\boldsymbol{x}, y) \in \mathcal{D}^{gen}}[l_{w}(\boldsymbol{x}, y)]$ in FL paradigm design, aim at learning $w$ that can capture cuer-invariant features with semantic consistency across clients and can perform well on general population data $\mathcal{D}^{gen}$ (\ie, the mixture of $\mathcal{D}_1, \mathcal{D}_2, \cdots,\mathcal{D}_K$).

\subsection{FedCSR Paradigm}
Our core idea is to extend knowledge distillation to FL to tackle the data heterogeneity issue. Different from previous KD tricks, we propose a mutual knowledge distillation (MKD) between local and global optimizations. To this end, two kinds of models are utilized to achieve communications between clients and server, \textit{i.e.}, linguistic model (parameterized by $\theta$) for textual information and visual model (parameterized by $w$) for visual information, respectively. Overall, the MKD method contains the following two aspects: (1) Visual-to-linguistic distillation (Vis2Lin). On the server side, a linguistic model is trained using the text data, where the global aggregation of local visual models is the teacher to guide the training of the linguistic model. (2) Linguistic-to-visual distillation (Lin2Vis). On the client side, $k$-th local client treats the linguistic model $\theta$ as the teacher to train a visual model $w_k$ using its own local data. MKD scheme can ensure that different clients learn a unified feature space, where different modalities (\textit{i.e.}, lip and hand) and linguistic dependencies are well aligned towards each other to enhance cross-modal semantic consistency.

\section{The Proposed Method}
Here, we will first introduce the architectures of the visual model and linguistic model. Then the proposed FedCSR framework with the MKD method will be described in detail, including visual-to-linguistic distillation (Vis2Lin) and linguistic-to-visual distillation (Lin2Vis). The overall framework is illustrated in Figure \ref{fig:2}.

\subsection{Model Architecture}
\begin{figure}[!t]
\centering
\includegraphics[width=0.9\linewidth]{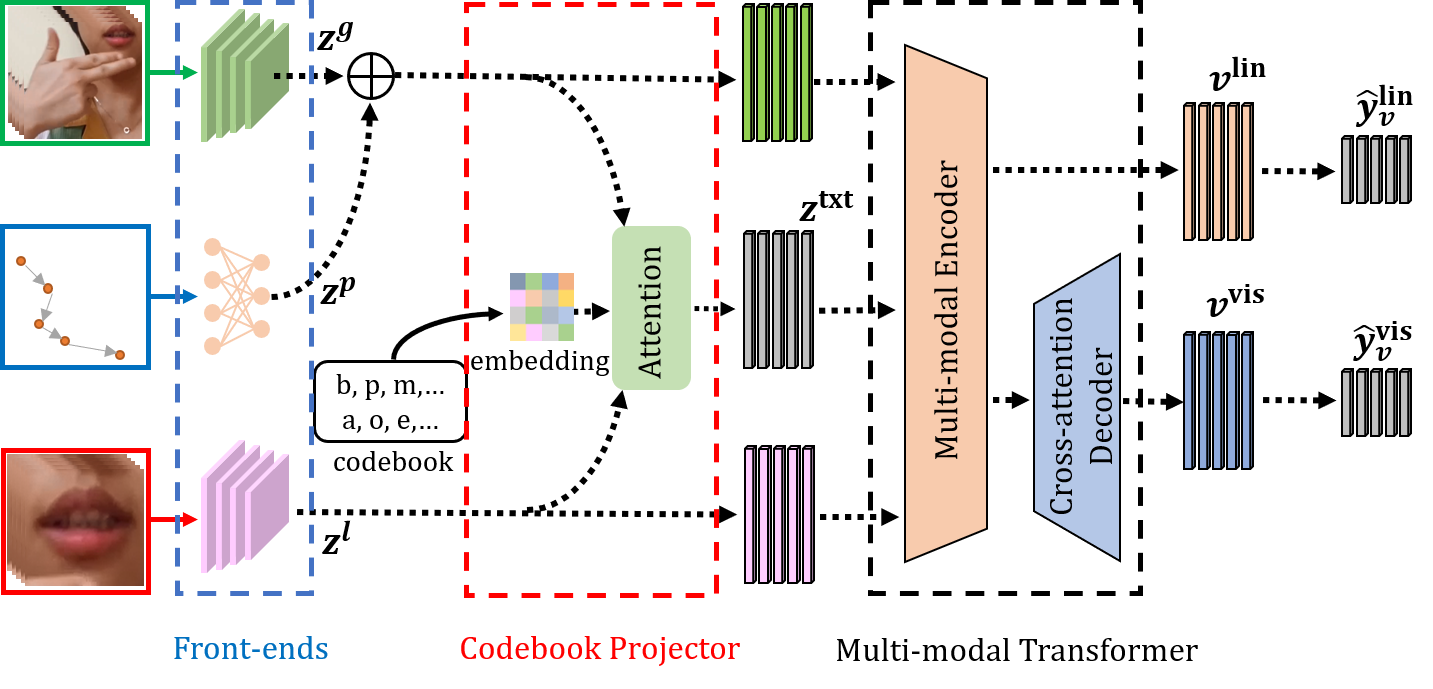}
\caption{Detailed architecture of the CMML model composed of visual front-ends, a codebook projector, and a multi-modal transformer. CMML produces two kinds of outputs $v^\text{lin}$ and $v^\text{vis}$ representing linguistic and visual features, respectively.}
\label{cmml}
\end{figure}

\balance 
\paragraph{Visual Model.} Motivated by the previous study \cite{liu2022cross}, we utilize the Cross-modal Mutual Learning (CMML) model as the visual model to capture visual information during local training. As shown in Figure \ref{cmml}, in the local visual model, a codebook projector is used to extract linguistic features of CS. The front-end adopts two CNNs and an MLP to extract the frame-wise features of the input video sequence, \textit{i.e.}, $\mathbf{z}^l, \mathbf{z}^g, \mathbf{z}^p \in \mathcal{R}^d$ for lip, hand shape, and hand position, where $d$ is the feature dimension. Then a following multi-modal transformer allows the free attention flows in the same self-attention layer for both linguistic and visual information. Finally, a cross-attention layer is used for the modality alignment between lip and hand guided by the shared linguistic features. Overall, the CMML model produces both visual ($v^\text{vis}$) and linguistic ($v^\text{lin}$)features for the final recognition task:
\begin{equation}
   v^\text{vis}, v^\text{lin} = \operatorname{CMML}(z^l, z^g, z^p; w).
\end{equation}

\paragraph{Linguistic Model.} An auto-encoder of time series data can support the learning of domain-invariant and modality-invariant linguistic features \cite{srivastava2015unsupervised}. Inspired by this, we train such a model on the server side using only textual data without access to image data on clients, it is called a linguistic model since it captures the linguistic properties of CS. In detail, the embedding layer of the linguistic model shares the same weights with the codebook projector of the visual model, which produces the hidden embeddings $z^{\text{txt}}$ of the text data $y$:
\begin{equation}
   z^{\text{txt}}=\operatorname{Projection}(\operatorname{ReLu}(\operatorname{Embedding}(y;\phi))),
   \label{eq:3}
\end{equation}
where $\phi$ is the weight of the embedding layer. Remaining part is a Seq2Seq model \cite{sutskever2014sequence} of Bi-LSTM \cite{huang2015bidirectional} structure parameterized by $\theta$ for variable-length temporal feature modeling:
\begin{equation}
   z^{\text{lin}}=\operatorname{Bi-LSTM}(z^{\text{txt}}, \theta),
   \label{eq:4}
\end{equation}
where $z^{\text{lin}}$ is the linguistic feature produced by the linguistic model. The output length is set in alignment with the visual model in each training iteration.

\subsection{Mutual Knowledge Distillation}

\subsubsection{Visual-to-Linguistic Distillation}
On the server side, FedCSR focuses on extracting the linguistic properties of the textual data, which should be aligned with the visual semantics of different local clients. To this end, we first consider learning a conditional distribution $Q: \mathcal{Y} \rightarrow \mathcal{X}$ to characterize such alignment, which is consistent with the ground-truth textual data distributions:
\begin{equation}
    Q=\underset{Q: \mathcal{Y} \rightarrow \mathcal{X}}{\arg \max } \mathbb{E}_{y \sim p(y)} \mathbb{E}_{x \sim Q(x \mid y)}[\log p(y \mid x)]
\label{v2l}
\end{equation}
where $p(y)$ and $p(y\mid x)$ are the ground-truth prior and posterior distributions of the target labels, respectively. We assume that $p(y)$ is known via available text sentences, while $p(y\mid x)$ is unknown due to unobserved local data. To make Equation (\ref{v2l}) optimizable \textit{w.r.t} $Q$, we utilize a linguistic model on the server side such that both input and output are the same text sequence from the global corpus $G=\{y_i\}_{i=1}^N$. Besides, directly optimizing Equation (\ref{v2l}) over the input space $ \mathcal{X}$ requires the user data profile with the risk of privacy leakage, which also brings computation costs on the input space of high dimension. A more approachable approach is to conduct a knowledge distillation to align features of the linguistic model with features of the visual model, which is more compact than the raw image data and can alleviate privacy concerns:

\begin{equation}
G^*=\underset{\theta,\phi}{\arg \min } \mathbb{E}_{y \sim p(y)} \left[ L_{KD}(  z^{\text{lin}}| v^{\text{lin}})\right],
\end{equation}
\begin{equation}
L_{KD}(  z^{\text{lin}}| v^{\text{lin}})=\frac{1}{2T}\sum_{i=1}^{T}\left\|z_i^{\text{lin}}-v_i^{\text{lin}}\right\|_2^2,
\end{equation}
where $T$ is the phoneme-level sentence length and $v^{\text{vis}}$ is the second last linguistic feature of the aggregated visual model. $\|\cdot\|_2$ denotes the $L$-2 norm of the feature vectors. Visual-to-linguistic distillation is presented to take the aggregated visual model as a teacher for optimizing the following objective:
\begin{equation}
   \mathcal{L}_{lin}=L(\hat{y}_z^{\text{lin}}, y) + \beta L_{KD}(z^{\text{lin}}| v^{\text{lin}}),
   \label{la}
\end{equation}
where $L$ is the classification loss such as cross-entropy loss, $\beta$ is a hyper-parameter. Given an arbitrary sentence $y$, optimizing Equation ($6$) only requires access to the embedding layers and predictor modules of users' visual models. Under the guidance of the aggregated visual models, the linguistic model can learn better text embeddings, which are consistent with the visual features from the aggregated visual models. In other words, the linguistic model can approximate a unified feature space with semantic consistency, which is aligned with the feature space of the visual model trained over the local data.

\subsubsection{Linguistic-to-Visual Distillation}
On the client side, the goal of local training is to train a visual model aligned with CS semantic knowledge, utilizing the linguistic model $\theta$ and the embedding $\phi$ dispatched from the server, which contains linguistic properties of textual corpus from the global view. However, it is difficult for different clients with heterogeneous data distributions to learn a unified feature space with consistent semantic information, since each client only has access to its own data $\mathcal{D}_k$. 

To alleviate this problem, we utilize the linguistic model as the teacher for training the visual model in each client, called linguistic-to-visual distillation, where visual models of different clients are constrained by the consistent linguistic information. During local training, a visual model takes multi-modal data (\textit{i.e.}, lip, and hand) as the input. The front-end and codebook projector in the CMML is used to extract visual $v^{\text{vis}}$ and linguistic $v^{\text{lin}}$ features, respectively.

\begin{algorithm}[!b]
	\caption{FedCSR Algorithm}
	\label{alg:algorithm1}
	\KwIn{Communication rounds $T$; Local epoch $M$; Global epoch $\tau$.} 
	
	Model initialization.
	
	\For{$t=0 \sim T-1$}{
            \tcp{local training:}
		clients download $\phi^{(t)},\theta^{(t)},w^{(t)}$ \\
		\ForEach{client $\mathcal{D}_i$}{
                \For{$m=0 \sim M-1$}{
    			    Compute $\mathcal{L}_{vis}$ as in Eq. \ref{lv}\\
                    $w_i^{(t)}\leftarrow w_i^{(t)}-\nabla\mathcal{L}_{vis}$
                }
            client uploads $\theta_i^{(t)}$ to server\\            
		}
		
            \tcp{Global training:}
            Model aggregation: $w^{(t+1)}=\frac{\sum_{i=1}^N w_i^{(t)}}{N}$;
            \For{$h=0 \sim \tau-1$}{
                Compute $\mathcal{L}_{lin}$ as in Eq. \ref{la}\\
                $(\theta^{(t)},\phi^{(t)})\leftarrow(\theta^{(t)},\phi^{(t)})-\nabla\mathcal{L}_{lin}$
            }
	}
	Return trained model parameters $\phi^{(T)},w^{(T)}$
\end{algorithm}

Specifically, both visual and linguistic features are aligned by the multi-modal self-attention with the cross-modal mutual learning strategy \cite{liu2022cross}, where the lip and hand features are aligned to the linguistic features of the codebook projector. Since the codebook projector shares the same embedding weights with the linguistic model, we additionally focus on the linguistic part ($v^{\text{vis}}$) produced by the CMML model parameterized by $w$. Given the input label sequence $y$, similar to Equation (\ref{eq:3}) and (\ref{eq:4}), $z^{\text{lin}}$ is the hidden feature vector obtained from the linguistic model. Then we distill linguistic knowledge of the linguistic model to the visual model using following objective:
\begin{equation}
G^*=\underset{w}{\arg \min } \mathbb{E}_{y \sim p(y)} \left[ L_{KD}(  v^{\text{lin}}| z^{\text{lin}})\right],
\end{equation}
Here, linguistic-to-visual distillation achieves semantic alignment between the local visual model and the linguistic model. Besides, the output logits $\hat{y}_z^{\text{lin}}$ of the linguistic model are also used as supervision for the sequential classification problem to assist the visual model to capture class-specific features in an earlier stage. In summary, besides the original supervision loss of the CMML model which is calculated on the visual model's two-way output, the objective of local training includes two extra parts, the supervision loss calculated on the linguistic model's output, and the knowledge distillation loss on the hidden feature of two models:
\begin{equation}
   \mathcal{L}_{vis}= L_{seq}(\hat{y}_v^\text{vis}, y) + L_{seq}(\hat{y}_v^{\text{lin}}, y) + \gamma L_{seq}(\hat{y}_z^{\text{lin}}, y) + \alpha L_{KD}(v^{\text{lin}}| z^{\text{lin}}),
   \label{lv}
\end{equation}
where $L_{seq}$ is the classification loss for the sequence, $\gamma$ and $\alpha$ are hyper-parameters. Since the model output and sentence label are not the same length, we use Connectionist Temporal Classification (CTC) loss \cite{graves2006connectionist} to align them. In this way, the visual models trained over the Non-IID visual CS data can learn a unified feature space consistent with the global linguistic model without accessing private data.

\subsection{FedCSR with MKD}
In fact, MKD is applied between visual and linguistic features in the FedCSR paradigm to fit the ACSR task. The training process is described in Algorithm \ref{alg:algorithm1}. In each communication round, the client downloads the linguistic and aggregated visual model from the server. Then, each client trains its local visual model on its own CS data, while the linguistic model is the teacher with the fixed weights. After local training, the clients upload the accumulated gradients and the server aggregates them via FedAvg. In global training, the linguistic model is trained on textual data, where the aggregated visual model is the teacher with fixed weights. Global and local training alternates in FL via MKD, thus the model eventually learns consistent semantic features for different clients.

\section{Experiments}
\subsection{Datasets}
The Chinese CS dataset \cite{liu2022cross} contains 4,000 CS videos corresponding to 4,000 Mandarin Chinese sentences, where 4 cuers in a total perform the Mandarin Chinese CS system to encode 1,000 Mandarin Chinese sentences. In detail, Chinese vowels and consonants are categorized into 40 phonemes, represented by a combination of 8 hand gestures and 5 hand positions with corresponding lip shapes. It should be mentioned that although French CS \cite{liu2018automatic} and British English CS \cite{multistream} are also used by researchers, French CS datasets contain very limited data for the single-cuer setting compared with Chinese CS dataset. The multi-cuer data of the British English CS dataset is not open-sourced. Thus, these two datasets are not included in this work. Only sentence-level labels are provided in the training and test set for sequence-to-sequence recognition, which is different from the most previous ACSR approaches requiring frame-level annotations. 

For FL experiments, video data produced by different cuers are assigned to different clients, hence creating a federated Non-IID scenario. For training and test sets splitting, 80\% of each cuer's data is stored on their client as local training data. The server collects the rest 20\% data from each client as the global test set. There is no sentence level overlapping between training and testing sets. The global corpus is also sentences extracted from the training set.

\subsection{Evaluation Metrics}
Our method evaluates the global model's performance on the test set using character error rate (CER) and word error rate (WER) to evaluate the recognition ability on phoneme and word levels. CER is calculated on phoneme level, \ie, each of the 40 phonemes is treated as a character. WER is more strict since it calculates errors on morphemes grouped by phonemes.

\subsection{Implementation Details}
 The ROIs of hand and lip are extracted from the original CS video, resized to $64 \times 64$ as the input. For the visual model, we use ResNet-18 \cite{he2016deep} as the front-end, where the first layer is replaced by a 3D convolutional layer with kernel size $5\times3\times3$. Hand position is a sequence of hand ROIs' coordinates, and the corresponding front-end network is a two-layer MLP. The multi-modal transformer has 3 encoder layers, where the rest of the structure is the same as \cite{liu2022cross}. Transformer parameters are randomly initialized, while the front-end is initialized by pre-trained weights on ImageNet. For the linguistic model, encoder and decoder are the Bi-LSTM structures with 2 and 4 hidden layers, respectively. The hidden state size is $512$. In local training, Adam optimizer is used with $\beta_1=0.9$, $\beta_2=0.98$, and $\epsilon=0.05$. The learning rate is set as 0.001. The training of the linguistic model also applies an Adam optimizer with $\beta_1=0.9$, $\beta_2=0.98$, $\epsilon=0.05$, and the learning rate is 0.0001. In each communication round, all clients participate in training and aggregation. The global training epochs per round are fixed as 10.

\subsection{Baselines}
To evaluate the effectiveness of the proposed approach, both previous centralized ACSR methods and several popular FL algorithms are adopted as the comparison methods, mainly including CNN + LSTM \cite{9287365}, CNN + CTC \cite{multistream}, JLF + COS + CTC \cite{wang2021cross}, Self-attention \cite{vaswani2017attention}, and CMML \cite{liu2022cross} as ACSR methods, as well as FedAvg \cite{mcmahan2017communication}, FedProx \cite{li2020federated}, and FedBN \cite{li2021fedbn} for FL baselines. The same model and parameter setting are applied for all comparisons.

\begin{table}[!t]
\centering
\small
\caption{The CER(\%) and WER(\%) results of ACSR task on Chinese CS dataset. FedCSR achieves significant improvement compared with previous methods.}
\begin{tabular}{lcc}
\hline
Method                                    & CER           & WER      \\ \hline
CNN + LSTM \cite{9287365}                 & 61.4          & 96.1     \\
CNN + CTC \cite{multistream}    & 41.9          & 83.4     \\
JLF + COS + CTC \cite{wang2021cross}      & 68.2          & 98.1      \\
Self-attention \cite{vaswani2017attention}& 38.8          & 78.6      \\
CMML \cite{liu2022cross}                  & 24.5          & 54.5      \\
\textbf{FedCSR (Ours)}                             & \textbf{14.8} & \textbf{39.5}   \\ \hline
\end{tabular}
\label{exp_comp}
\end{table}

\begin{table}[!t]
\centering
\small
\caption{Comparison results with FL baselines. FedCSR achieves the best performance in all local epochs settings.}
\begin{tabular}{l|cc|cc|cc}
\hline
Local epochs   & \multicolumn{2}{c|}{1}         & \multicolumn{2}{c|}{3}         & \multicolumn{2}{c}{5}       \\ \hline
FL Method              & CER           & WER           & CER           & WER           & CER         & WER           \\ \hline
FedAvg \cite{mcmahan2017communication}        & 17.8          & 46.8          & 20.0            & 49.6          & 20.5        & 50            \\
FedProx \cite{li2020federated}       & 17.1          & 44.3          & 19.7          & 50.8          & 20.4        & 51.8          \\
FedBN \cite{li2021fedbn}       &18.1          & 47.3          & 17.9          & 45.7          & 21.7        & 53.3          \\
\textbf{FedCSR (Ours)} & \textbf{14.8} & \textbf{39.5} & \textbf{17.3} & \textbf{45.7} & \textbf{18.0} & \textbf{46.5} \\ \hline
\end{tabular}
\label{exp_fed}
\end{table}

\begin{figure}[!t]
    \centering
    \includegraphics[width=0.9\linewidth]{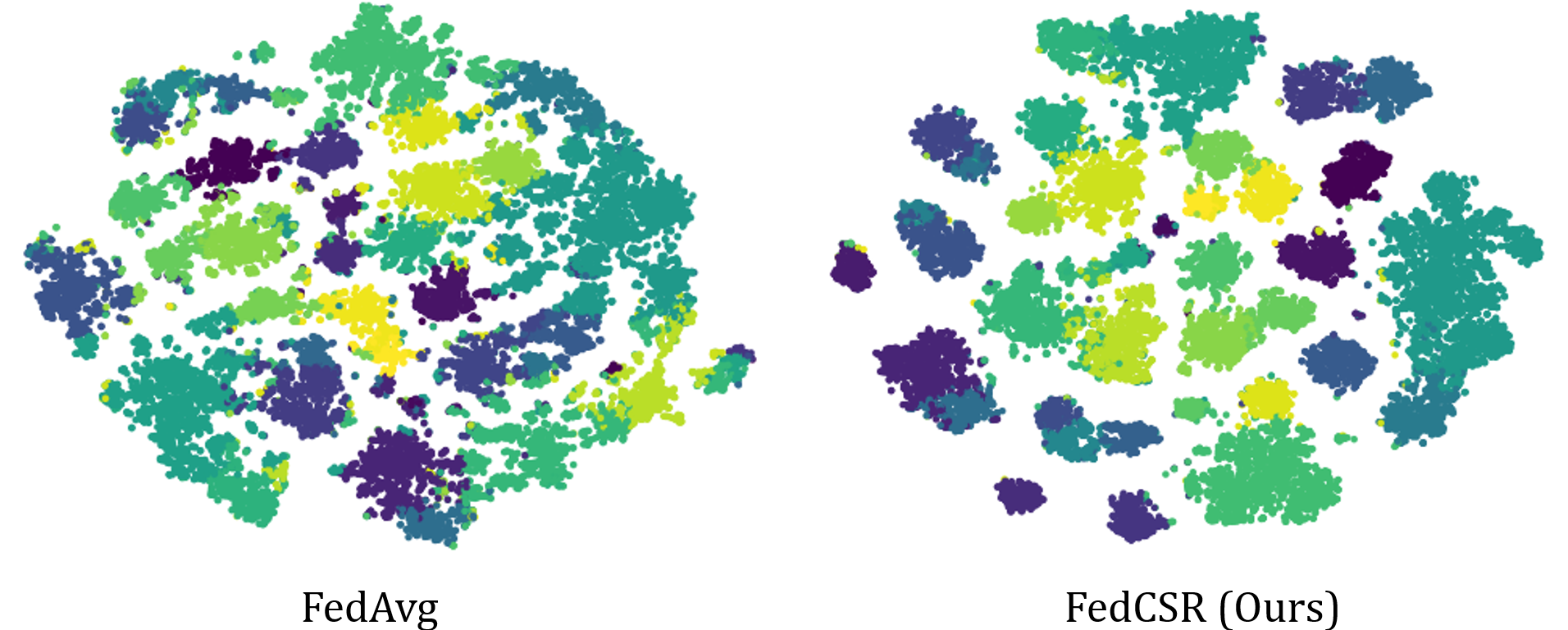}
    \caption{T-SNE visualization of hidden linguistic vector in phoneme level. The clusters represent the distribution of hidden linguistic vectors for 40 phonemes in Chinese CS (colors are reused). FedCSR method produces clearer clusters of hidden vector.}
    \label{tsne}
\end{figure}

\begin{figure*}[!t]
    \centering
    \includegraphics[width=0.85\linewidth]{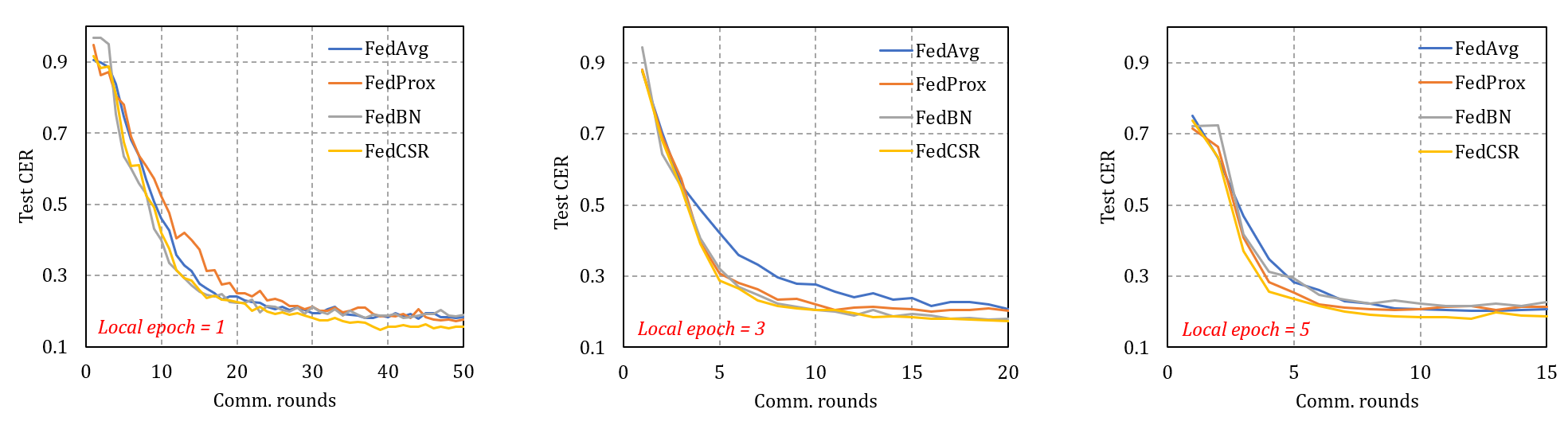}
    \caption{Effects of the communication rounds. FedCSR outperforms other FL methods in performance and convergence speed.}
    \label{traincurv}
\end{figure*}

\begin{figure*}[!t]
    \centering
    \includegraphics[width=0.85\linewidth]{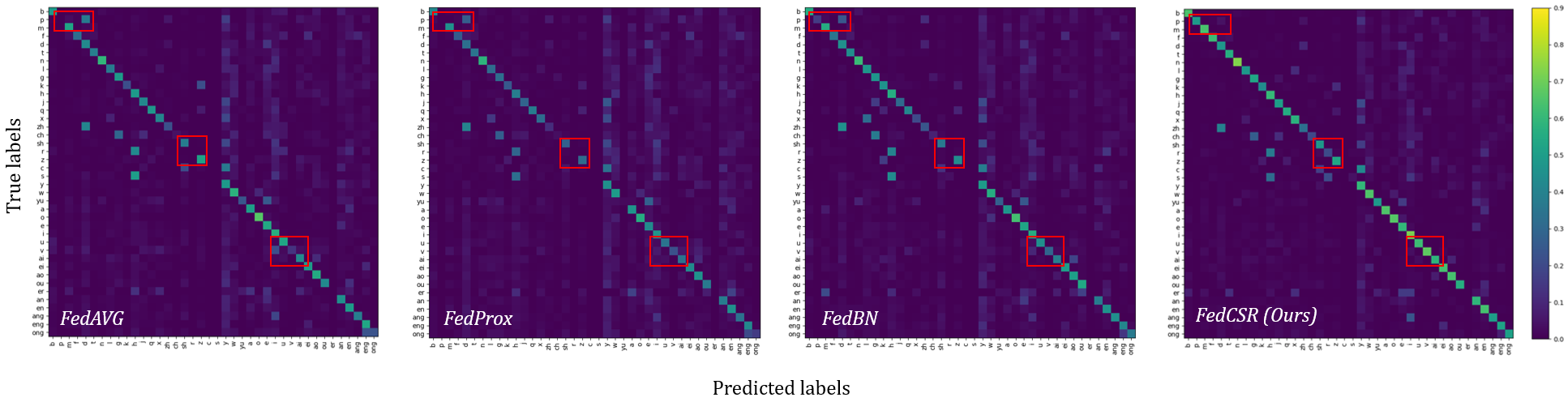}
    \caption{Phoneme level confusion matrix of FedCSR and other FL methods. FedCSR outperforms in hard sample classification.}
    \label{confusion}
\end{figure*}

\subsection{Comparison Results}
\subsubsection{Compared with previous ACSR Methods} As shown in Table \ref{exp_comp}, our method achieves SOTA performance of 14.8\% CER and 39.5\% WER, outperforming previous centralized training SOTA \cite{liu2022cross} by 9.7\% lower CER and 15.0\% lower WER, while our method can well protect the data privacy of each cuer on the clients. The main reason lies that the proposed MKD strategy can better align the visual and linguistic information, thus achieving a semantic consistency between different modalities.

\subsubsection{Compared with classical FL Methods} Table \ref{exp_fed} shows comparison results of FL algorithms. Our model is trained under 1, 3 and 5 local epochs setting. The best performance is obtained when the local training epoch is 1. Since the data heterogeneous issue is significant for different cuers on different clients, there may be severe domain drift when the local visual model is trained for several rounds with multiple local epochs. Besides, the proposed MKD strategy can achieve global semantic consistency for multi-modal data of different cuers, and thus can achieve higher recognition accuracy than other FL algorithms.

\begin{table}[!h]
\centering
\small
\caption{Ablation studies for different distillation strategies, under 3 local epochs setting.}
\begin{tabular}{ccccc}
\hline
shared embedding & distill visual. & distill ling. & CER           & WER           \\ \hline
\XSolidBrush             & \XSolidBrush                & \XSolidBrush              & 20.0          & 49.6          \\
\Checkmark               & \XSolidBrush                & \XSolidBrush              & 19.7          & 50.0          \\
\XSolidBrush             & \XSolidBrush                & \Checkmark                & 19.0          & 48.2          \\
\Checkmark               & \Checkmark                  & \XSolidBrush              & 18.1          & 47.7          \\
\Checkmark               & \XSolidBrush                & \Checkmark                & \textbf{17.3} & \textbf{45.7} \\ \hline
\end{tabular}
\label{exp_abn}
\end{table}

\begin{table}[!h]
\centering
\small
\caption{Ablation studies for hyper-parameters in mutual knowledge distillation, under 3 local epochs setting.}
\begin{tabular}{ccccc}
\hline
$\alpha$ & $\beta$ & $\gamma$ & CER           & WER           \\ \hline
0     & 0    & 0     & 20            & 49.6          \\
0.005   & 0    & 0.5   & 19.7          & 50.3          \\
0.005   & 0.005  & 0     & 18.8          & 47.3          \\
0     & 0    & 0.5   & 18.3          & 47.1          \\
0.005   & 0.005  & 0.5   & \textbf{17.3} & \textbf{45.7} \\ \hline
\end{tabular}
\label{exp_loss}
\end{table}

\begin{table*}[!t]
\centering
\small
\caption{Results of leave-one-domain-out experiment, with local epoch as 3. HS, LF, WT and XP are four cuers.}
\begin{tabular}{l|cc|cc|cc|cc|cc}
\hline
Test Cuer       & \multicolumn{2}{c|}{HS}       & \multicolumn{2}{c|}{LF}       & \multicolumn{2}{c|}{WT}       & \multicolumn{2}{c|}{XP}       & \multicolumn{2}{c}{Avg.}      \\ \hline
FL Method              & CER           & WER           & CER           & WER           & CER           & WER           & CER           & WER           & CER           & WER           \\ \hline
FedAVG        & 33.7          & 72.5          & 34.5          & 73.7          & 37.3          & 79.6          & 43.8          & 90.3          & 37.3          & 79.0          \\
FedProx       & 31.6          & 67.7          & 32.2          & 71.4          & 36.1          & 82.7          & 39.3          & 81.6          & 34.8          & 75.9          \\
FedBN         & 30.7          & 67.6          & 32.8          & 71.0          & 36.3          & 81.2          & 38.7          & 81.2          & 34.6          & 75.3          \\
\textbf{FedCSR (Ours)} & \textbf{27.7} & \textbf{65.4} & \textbf{30.9} & \textbf{68.9} & \textbf{34.2} & \textbf{77.2} & \textbf{37.0} & \textbf{79.8} & \textbf{32.6} & \textbf{72.8} \\ \hline
\end{tabular}
\label{exp_ood}
\end{table*}

\vspace{-0.2cm}

\subsection{Visualizations}
\subsubsection{Feature Distribution.} As shown in Figure \ref{tsne}, we utilize the t-SNE \cite{van2008visualizing} technique to visualize the distribution of the learned linguistic feature $v^{\text{lin}}$ in the visual model on phoneme level. Different colors of the dots represents different phonemes in Chinese CS. It is observed that the feature distribution of $v^{\text{lin}}$ is significantly correlated with different phoneme classes, where our method can produce more discriminative clusters of hidden vectors than the model trained by FedAvg. Thus, the proposed MKD approach can effectively achieve the alignment for cross-modal visual features with the textual label.

\subsubsection{CER Curve} Figure \ref{traincurv} shows the decreasing of test CER along with a number of communication rounds. Results show that due to data heterogeneity, the convergence speed of FedAVG is slower than our method. Although FedProx and FedBN decrease as fast as our method in the beginning, the curve converges in an earlier stage. In all local epoch settings, our method shows superiority in both performance and convergence speed, which indicates our MKD method is stable in the federated training process.

\subsubsection{Confusion Matrix} The prediction of the model can be decoded into a sequence of phonemes in a greedy approach. By aligning decoded sentences with ground truth labels, we draw the confusion matrix on the phoneme level. As shown in Figure \ref{confusion}, the colors of diagonal elements show correctly classified phonemes. Generally, consonants that share the same hand posture are hard to classify for all models, as well as some compound vowels since lip shapes are usually ambiguous and vary with different cuers. Our method shows better performance in hard sample classification, \eg, consonants such as \textit{`p'}, \textit{`r'} and vowels such as \textit{`v'}, \textit{`ong'}, which indicates that our method successfully extracted linguistic information in 
global training that assists the visual model to distinguish ambiguous CS phonemes through MKD.

\subsection{Ablation Studies}
\subsubsection{Component Analysis} We conduct an ablation study on each module of our proposed method. In this part, we set the local epoch as $3$ for all experiments. Results are shown in Table \ref{exp_abn}. 

(1) \textit{Shared embedding} means whether sharing the weights between the embedding layer of the linguistic model and the codebook projector of the visual model. Results show that shared embedding does not improve model performance individually, but it yields a significant performance improvement of $1.7\%$ CER when combined with MKD. This indicates that the embedding is not only required to be unified but also needs an effective training process. In our MKD process, the shared embedding is trained using textual data, such that embedded vectors contains the linguistic properties CS phonemes, which gradually assists phoneme level prediction.

(2) The CMML model produces two kinds of hidden vectors $v^{\text{lin}}$ and $v^{\text{vis}}$. Following the insight provided by \cite{liu2022cross}, $v^{\text{lin}}$ contains more linguistic information where as $v^{\text{vis}}$ is the decoded visual feature. We try both for our mutual knowledge information. \textit{distill visual.} and \textit{distill ling.} respectively denotes that we align $v^{\text{vis}}$ and $v^{\text{lin}}$ with $z^{\text{lin}}$ for semantic consistency. Results show that distilling $v^{\text{lin}}$ is the best strategy, where the proposed MKD strategy successfully transfers global semantic knowledge into the visual model through $v^{\text{lin}}$, and the decoded visual feature $v^{\text{vis}}$ is only responsible for alignment with ground truth label.

\subsubsection{Hyper-parameter Analysis} We conduct further ablation studies on hyper-parameters of loss terms in MKD. In the learning process, except for the CTC loss for output sequence, two loss terms assist the local model to learn linguistic features in latent space. Respectively, we use KD loss for semantic feature alignment, with term coefficients $\alpha$ and $\beta$ respectively for local and global training, and CTC loss for semantic model output with coefficient $\gamma$ for vision back-bone supervision. Remarkable results are shown in Table \ref{exp_loss}. Compared with FedAvg ($\alpha = \beta = \gamma = 0$), both components improve model performance independently, and combining all loss terms gives the best performance. Results of single linguistic-to-visual distillation ($\alpha = 0.005, \beta = 0, \gamma = 0.5$) do not bring improvements, which indicates both training directions are crucial for visual-linguistic MKD to be valid.

\subsection{Out-of-Distribution Results}
To further examine the effectiveness of our approach, we use the leave-one-domain-out strategy as a complementary experimental setting. Specifically, in our 4-client FL paradigm, we use 3 clients for training and leave the rest one client as the test set. During training, the test set is an unseen domain with out-of-distribution (OOD) data. Results in Table \ref{exp_ood} show a performance drop compared with previous results, which indicates Non-IID feature of CS data by different cuers significantly impacts model performance. Compared with FL baselines, our model achieves the best performance on all 4 OOD test sets and the best overall performance.

\section{Conclusion}
In this work, we consider the FL setting for the ACSR task to mitigate privacy concerns. To this end, we propose a new framework called FedCSR to train an ACSR model over the decentralized CS data without sharing users' data. In particular, we propose a mutual knowledge distillation method to achieve cross-modal semantic consistency. The server learns a linguistic model guided by the aggregation of the local visual models via visual-to-linguistic distillation. In local training, the visual model of each client is trained with the linguistic-to-visual distillation with the linguistic model as the teacher. The experimental results on the Chinese CS dataset with multiple cuers demonstrate that our method successfully learns a unified feature space of the linguistic and visual information, which significantly improves the recognition performance compared with mainstream FL baselines and previous centralized SOTA ACSR methods. Future work will further explore the domain generalization in ACSR under the FL setting, eventually filling the gap between normal-hearing and hearing-impaired people.

\section{Acknowledgements}
This work was supported by grants from the National Natural Science Foundation of China (No. 62101351) and the GuangDong Basic and Applied Basic Research Foundation
(No.2020A1515110376).

\bibliographystyle{ACM-Reference-Format}
\bibliography{sample-base}

\end{document}